\documentstyle[amssymb,epsfig]{l-aa}
\begin{document}
\thesaurus{10.08.1, 12.07.1, 12.04.1, 10.07.2}
\title{Gravitational Microlensing by Globular Clusters}
\author{Philippe Jetzer\inst{1,2}, Marcus Str\"assle\inst{2}
and Ursula Wandeler\inst{2}}
\offprints{marcus@physik.unizh.ch}
\institute{
Paul Scherrer Institut, Laboratory for Astrophysics,
CH-5232 Villigen PSI
\and
Institut f\"ur Theoretische Physik
           der Universit\"at Z\"urich,
           Winterthurerstrasse 190,
           CH-8057 Z\"urich}

\date{Received; accepted}

\maketitle
\markboth{Gravitational Microlensing by Globular Clusters}{Gravitational
Microlensing by Globular Clusters}
\begin{abstract}
Stars in globular clusters can act either as sources for MACHOs 
(Massive Astrophysical Compact Halo Objects) located
along the line of sight or as lenses for more distant background stars.
Although the expected rate of microlensing events is small,
such observations can lead to very
useful results. In particular, one could get information on the shape of the
galactic halo along different lines of sight, allowing to better constrain
its total dark matter content.

Moreover, on can also infer the
total dark matter content of globular clusters, which is presently not well
known. To this latter purpose, 
we analyse the microlensing events towards the galactic bulge,
which lie close to the three globular 
clusters NGC 6522, NGC 6528 and NGC 6540.
We find evidence that some microlensing events are indeed due to MACHOs
located in the globular clusters, suggesting, therefore, that these clusters
contain a significant amount of dark matter.
\end{abstract}

\section{Introduction}

An important problem in astrophysics concerns the nature of the dark
matter in galactic halos, whose presence is implied by the observed
flat rotation curves in spiral galaxies. 
Microlensing (Paczy\'nski \cite{Pac86}) allows the detection of MACHOs
in the mass range $10^{-7}<
M/M_{\odot}<1$ (De R\'ujula, Jetzer, Mass\'o \cite{drj}) in the halo, disk or bulge 
of our galaxy. Till now, more than
$15$ microlensing events have been found towards the Large Magellanic Cloud
(LMC) (Alcock, Akerlof, Allsman et al. \cite{Mac93}, Alcock, Allsman, Alves
et al., \cite{Alc97b}, Auburg, Bareyre, Br\'ehin et al \cite{Ero93}),
one event towards the Small Magellanic Cloud (SMC) (Alcock, Allsman, Alves et al.
\cite{Alc97c}, Palanque-Delabrouille, Afonso, Albert et al. \cite{npd},
Udalski, Kubiak \& Szyma\'nski \cite{Uda97}) and about
$200$ events towards the galactic bulge (Alcock, Allsman, Alves et al. \cite{Alc97a},
Udalski, Szyma\'nski, Stanek et al. \cite{Uda94},  Alard \& Guibert \cite{Duo97}).

However, in spite of the many events, several questions are still open, in
particular on the mass and the location of the lenses. In fact, from the
duration of a single microlensing event, one cannot infer directly the mass
of the lens, since its distance and transverse velocity are generally not
known.
To break this degeneracy it has been proposed to perform parallax
measurements (Gould \cite{Gou97}), which however require the use of 
space satellites.

Globular clusters could be in many respect very useful to solve some
of these problems. In fact, microlensing searches using globular clusters as 
targets could probe different lines of sight in addition to the ones towards 
the LMC or the SMC, this way allowing to better determine the spatial
distribution of the MACHOs (Jetzer \cite{Jet91}).
Since in globular clusters much less stars than compared to the LMC or SMC
can be used as targets, one would have to monitor many globular clusters
in order to get some microlensing events. Gyuk \& Holder (\cite{gyu}) and
Rhoads \& Malhotra (\cite{rhoa}) have studied this possibility and shown 
that this way interesting
galactic structure information can be extracted allowing to distinguish
between different halo models.

Another possibility is to search for microlensing of background stars by
MACHOs located in foreground globular clusters. Such an observation can 
in addition give
important information on the total mass of globular clusters.
It has been argued recently that 
a large fraction of their mass (around $50$\%) is dark, 
which could be in the form of brown dwarfs, low-mass stars
or white dwarfs (Heggie et al. \cite{heg}, Taillet, Salati \& Longaretti \cite{tail},
\cite{tail2}). 
Moreover, one expects that the heavy stars tend to sink
towards the cluster cores, whereas the light objects populate the outskirts.
Hence, the dark component of the cluster is not similarly concentrated
towards the center as the bright stars which eases observation. 

The idea --as originally proposed by Paczy\'nski
(\protect\cite{Pac94})-- is to monitor globular clusters like 47 Tuc or
M22 in front of the rich background of either the SMC or the galactic bulge. 
In this case, when the lens belongs to the 
cluster population, its distance and velocity are roughly known. The
velocity is defined by the dispersion velocity of the cluster stars 
together with the overall transverse velocity of the cluster as a whole.
Knowing approximately the distance and the velocity of the lens would allow
to extract from a microlensing event the mass of the lens with an accuracy
of $\sim$ 30\%.

Due to these reasons, it is important to study in more detail
microlensing by globular clusters either using their stars as sources,
or the dark matter contained in them as lenses for more distant stars 
(Wandeler \cite{wand}).
In this paper we discuss both aspects in detail. 
Although the mass
distribution of the luminous part of the cluster, as inferred from the
observation of the distribution of the red giant population, agrees well with
a King model, we do not consider this to be representative for the
population of light objects. Taillet, Longaretti \& Salati (\cite{tail}, \cite{tail2}) have shown that
in an isolated globular cluster thermalisation between the different
populations does occur. However, globular clusters --especially the ones towards
the bulge-- tidally interact with the surrounding material which
might counteract thermalisation, hence the
consideration of alternative mass distributions should be taken into
account and variations of the microlensing event rate due to it might give 
some hints at the dynamical history of the cluster. 

We also analyse the
microlensing events towards the galactic bulge, which are close to the three
globular clusters NGC 6522, NGC 6528 and NGC 6540. These clusters lie
within the observation fields of the MACHO and OGLE teams.
We find evidence that some microlensing events are indeed due to MACHOs   
located in the globular clusters, suggesting therefore that these clusters
contain a significant amount of dark matter.

The paper is organized as follows: in Sect.~2 we introduce briefly
the basics of microlensing. In Sect.~3 we discuss as an example
the globular cluster 47 Tuc, which will then be used to estimate 
in a very conservative way the optical 
depth and the lensing rate for other clusters as well.
In Sect.~4 we present microlensing using globular clusters towards
the galactic bulge. In particular, we analyse the events as reported
by the MACHO and the OGLE collaboration lying within a distance of $30\,$pc
around the centers of NGC 6522, NGC 6528 and NGC 6540. In Sect.~5 we 
conclude with a short summary of our results. 

\section{Basics of microlensing}

For completeness we give here a short summary of the
most important formulae of gravitational microlensing; 
for more details see for instance Jetzer (\cite{Jet97}).

The time dependent magnification of a light source due to gravitational 
microlensing is given by 
\begin{equation}\label{pj5}
A(t)=\frac{u_{\circ}^2+t^2/T_{\circ}^2+2}{\sqrt{(u_{\circ}^2+t^2/T_{\circ}^2)(u_{\circ}^2+
t^2/T_{\circ}^2)+4}}\,,
\end{equation}
with
$u_{\circ}=d_{min}/R_E$, where $d_{min}$ is the minimal distance of the MACHO
from the 
line of sight between the source and the observer. $R_E$ is the 
Einstein radius, defined as
\begin{equation}\label{pj2}
R_E^2=\frac{4GMD}{c^2}x(1-x)\,,
\end{equation}
with $x=s/D$, $D$ and $s$ are the distances to the source and  
the MACHO, respectively. 
$v_T$ is the relative transverse velocity of the involved objects.
$T_{\circ}=R_E/v_T$ is the characteristic time for the lens to travel the
distance $R_E$.

The probability $\tau_{opt}$, that a source is found within a radius 
$R_E$ of some MACHO, 
is defined as 
\begin{equation}\label{pj3}
\tau_{opt}=\int_{0}^{1}\frac{4\pi G}{c^2}\rho(x)D^2x(1-x)\,dx\,,
\end{equation}
with $\rho(x)$ the mass density of microlensing matter at the distance 
$s=xD$ from us along the line of sight. 

The microlensing event rate is given by (De R\'ujula, Jetzer, Mass\'o \cite{Jet90}, 
Griest, Alcock, Axelrod et al. \cite{grie})
\begin{equation}\label{l1}
\Gamma = 2u_{TH} D\int R_E(x)v_Tf_T(v_T)\frac{dn}{d\mu}dv_Tdxd\mu\,,
\end{equation}
where $\mu=M/M_{\odot}$ and $f_T(v_T)$ is the transverse velocity
distribution.
The maximal impact parameter $u_{TH}$ is related to 
the threshold magnification $A_{TH}$ by Eq.~(\ref{pj5}).
In the following we take $A_{TH}=1.34$ which corresponds to $u_{TH}=1$.
$\frac{dn(x)}{d\mu}$ is the MACHO number density, which for a spherical
halo is given by
\begin{equation}\label{MWden}
\frac{dn(x)}{d\mu}=\frac{dn_{\circ}}{d\mu}
\frac{a^2 +R_{GC}^2}{a^2+R_{GC}^2+D^2x^2-2DR_{GC}\cos\alpha}\,,
\end{equation}
here $\alpha$ denotes the angle between the line of sight and the direction
towards the galactic center. $n_{\circ}$ is assumed not to depend on $x$ and is
normalized such that
\begin{equation}\label{norma}
M_{\odot}\int \mu \,d\mu\frac{dn_{\circ}}{d\mu}=\rho_{\circ}\simeq 7.9\times
10^{-3}\,\frac{M_{\odot}}{{\rm pc}^3}
\end{equation}
equals the local dark matter mass density. $R_{GC}\simeq 8.5\,{\rm kpc}$
is the distance from the Sun to the galactic center and 
$a\simeq 5.6\,{\rm kpc}$ is the core radius of the halo.
 
For the average lensing duration one gets the following 
relation
\begin{equation}\label{l2}
\langle T\rangle = \frac{2\tau_{opt}}{\pi\Gamma}u_{TH}\,.
\end{equation}

\section{The system SMC-47 Tuc: a paradigm for cluster lensing}

In this section we thoroughly discuss the basics of microlensing by a 
globular cluster. We use the system SMC-47 Tuc as an example, but the
results, by appropriately scaling them, are valid for other globular 
clusters as well.

First, we discuss simple models of 47 Tuc which will then be used for
the computation of the optical depth and the microlensing event rate
for different geometries of lens and source.

The globular cluster 47 Tuc (NGC 104) lies at galactic coordinates $l=305.9^{\circ}$, 
$b=-44.89^{\circ}$. For the distance we assume $4.1\,$kpc, although in the
literature values up to $4.7\,$kpc are quoted (Harris \cite{harr}). 
Our choice will rather underestimate the optical depth. 
The position is such that it overlaps with a part of the outer 
region of the SMC, which makes it an interesting object. 
Globular clusters are small objects compared to the scale of their distance, 
hence they are well suited for gravitational lensing, 
since one may assume the distance of their stars to be the same 
for all practical purposes.

\subsection{Spatial density and velocity dispersion for 47 Tuc}

For the calculation of the lensing rate 
we need to know the spatial distribution of the dark matter
in the globular cluster. Since this is not known, 
we will instead discuss models for the total mass of the cluster. 
To get the mass density $\rho _d$ of the dark objects, 
we then make the simplifying assumption
that $\rho _d$ is proportional to the total mass density $\rho$, i.e.
$\rho _d$ is given by 
\begin{equation}\label{lensden}
\rho _d = f\,\rho = \frac{M_{dark}}{M_{tot}} \rho\,,
\end{equation}
where $M_{dark}$ is the total mass of the MACHOs in the cluster. We are
aware, that this assumption is oversimplified, however, as long
as the content of dark matter in globular clusters is not known, it is
one way to parametrize our ignorance. 
Moreover, since the expected event
rate for 47 Tuc is about one event per year or even less, we do not consider
multi-mass models for 47 Tuc. We postpone the discussion of them to Sect.~4,
when we discuss lensing by globular clusters towards the bulge in which
situation the model can be tested due to the higher number
of microlensing events.

The simplest model of a globular cluster is a
self-gravitating isothermal sphere of identical
"particles" (stars). The equilibrium distribution function in phase-space 
coordinates is
\begin{equation} \label{fdist}
f(\vec x,\vec v) = \frac{\rho_{\circ}}{\left(2 \pi \frac{k_B T}{m}
\right)^{3/2}} \ \exp \left(\frac{m(\Phi(\vec x) - \frac{1}{2} v^2)}{k_B T}
\right)\,,
\end{equation}
where $\Phi(\vec x)$ is the gravitational potential of the cluster, $T$ the
temperature, $k_{\rm B}$ Boltzmann's constant and $m$ the mass of a
"particle". Taking into account the spherical symmetry of the globular cluster
and defining the one-dimensional velocity dispersion $\sigma$ as
\begin{equation}
\sigma  = \sqrt{\frac{k_B T}{m}}\,,
\end{equation}
Eq.~(\ref{fdist}) reads
\begin{equation} \label{fdistr}
f(r,\vec v) = \frac{\rho_{\circ}}{(2 \pi \sigma ^2)^{3/2}} \ \exp
\left(\frac{\Phi(r) - \frac{1}{2} v^2}{\sigma ^2} \right)\,.
\end{equation}
Here $r$ is the radial distance relative to the cluster center.
Integrating over all velocities, we get
\begin {equation} \label{expphi}
\rho (r) = \rho_{\circ}  \exp \left(\frac{\Phi(r)}{\sigma ^2} \right)\,.
\end{equation}
Inserting Eq.~(\protect\ref{expphi}) into the Poisson
equation for the gravitational potential, one obtains
\begin{equation} \label{poisson}
\frac{d}{dr} \left(r^2 \ \frac{d \ln \rho}{dr} \right) = - \frac{4 \pi G}{\sigma
^2}  r^2  \rho\,.
\end{equation}
A solution of this equation is
\begin{equation} \label{singd}
\rho (r) = \frac{\sigma ^2}{2 \pi G r^2}\,.
\end{equation}
The mass density given in Eq.~(\ref{singd}), together with the integrated 
velocity distribution Eq.~(\ref{fdistr}), defines the singular isothermal 
sphere.

To avoid the singularity at the origin we rescale the variables 
($\tilde r = r/r_c,\,\tilde \rho = \rho/\rho_{\circ}$) and find
a non-singular solution which can be
approximated for $\tilde r < 2$ by (for details see Binney \& Tremaine \cite{Bin87})
\begin{equation} \label{regd}
\tilde \rho(\tilde r) = \frac{1}{(1 + \tilde r^2)^{3/2}} \quad
\mbox{for} \ \tilde r < 2  
\end{equation}
and for $\tilde r >> 2$ with the singular function given in Eq.~(\ref{singd}).

An observable quantity, connected with the surface density of stars in
a cluster, is the surface brightness. The luminosity function
$\Phi (M)$ gives the relative number of stars with absolute magnitude $M$ in
the range [$M-1/2$, $M+1/2$]. The number surface density of the stars can 
be computed from the observed surface brightness. Assuming that all stars in the
cluster have the same mass, we can compare the above densities with surface
brightness measurements. The mass surface density $\Sigma$ is 
derived from the mass
density by the integration 
\begin{equation}
\Sigma (b) = \int_{-\infty}^{\infty} \rho (\sqrt{l^2 + b^2}) dl \ ,
\end{equation}
where $b$ is the projected radial distance.
For the density of the singular isothermal sphere, given by Eq.~(\ref{singd}),
the surface
density is
\begin{equation} \label{singsd}
\Sigma (b) = \frac{\sigma ^2}{2 G b} \,.
\end{equation}
Similarly, the density $\tilde\rho(\tilde r)$in Eq.~(\ref{regd}) leads to
the surface density
\begin{equation} \label{regsd}
\Sigma (b) \propto \frac{2}{1 + \tilde b^2} \,,
\end{equation}
where $\tilde b = b/r_c$.

Since Eq.~(\ref{regd}) fits the regular solution of Eq.~(\ref{poisson}) in the
range $\tilde r < 2$, we find that the corresponding surface density 
falls
to roughly half of its central value at the core radius $r_c$. 
Knowing $r_c$ and $\sigma$ from observations, the central
density $\rho _{\circ}$ is determined in the isothermal model by
\begin{equation}\label{corerad}
r_c=\sqrt{\frac{9\sigma^2}{4\pi G\rho_{\circ}}}\,.
\end{equation}
With $r_c=0.52\,$pc as given in Lang (\cite{Lang:92}) and a velocity dispersion of 
$\sigma=10\,{\rm km/s}$
(Binney \& Tremaine \cite{Bin87}), 
Eq.~(\ref{corerad}) for 47 Tuc yields
$
\rho _{\circ} = 6.0 \times 10^4 \,{\rm M}_{\odot}/{\rm pc}^3\,.
$

King (\cite{King:62}) found that the above 
surface density functions (\ref{singsd}), (\ref{regsd}) fit well 
star counts up to a limiting radius
$r_{t}$
(called tidal radius), where the measured surface density drops sharply to 
zero. Thus a
better fit to the surface density of the cluster
is given by
\begin{equation}
\Sigma (b) \propto \left(\frac{1}{\left[1+(b/r_c)^2
\right]^{1/2}} - \frac{1}{\left[1+(r_t/r_c)^2
\right]^{1/2}} \right)^2 \,.
\end{equation}
With a spherical symmetric density
\begin{equation}
\rho (r) = -\frac{1}{\pi} \int_r^{r_t} \left[\frac{d}{db} \Sigma(b) 
\right] \frac{db}{(b^2-r^2)^{1/2}} \,,
\end{equation}
the corresponding mass density becomes
\begin{eqnarray} \label{Kingd}
\rho (r) = \frac{k}{\pi r_{c} [1 + (r_{t}/r_{c})^{2}]^{3/2}} 
\frac{1}{z^{2}} \left[\frac{1}{z} \arccos{z} - (1 - z^{2})^{1/2} \right]\,, & &
\nonumber \\
& & 
\end{eqnarray}
where
$$ z = \left[\frac{1 + (r/r_{c})^{2}}{1 + (r_{t}/r_{c})^{2}}
\right]^{1/2} \ $$
and $k$ is chosen such that $\rho (0) = \rho_{\circ}$ (King \cite
{King:62}). 
Integrating this density with a tidal radius 
$r_{t} = 60.3\,{\rm pc}$
(Lang \cite{Lang:92}), one finds a total mass of $M_{{\rm tot}}^{{\rm King}} = 
 3.5 \times 10^{5} \,M_{\odot}$ for 47 Tuc.

\begin{figure}
\leavevmode
\epsfig{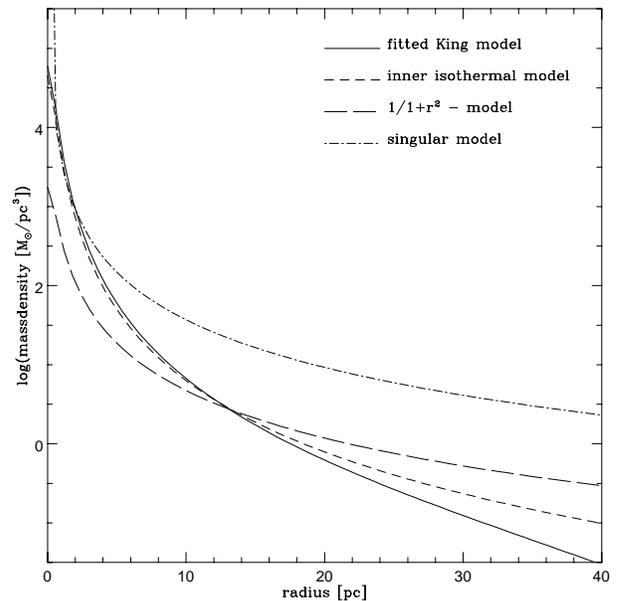}
\caption{The four different mass density models used in the calculations,
plotted as a function of the radial distance $r$. For clarity the plot ends at 
$r=40\,{\rm pc}$ rather than at $r=r_t=60.3\,{\rm pc}$.}
\label{massden}
\end{figure}

To study the dependence of the optical
depth and the lensing rate on different mass distributions, we will use the
following four
models (see
Fig.~\ref{massden}). 
\begin{description}
\item[1. Fitted King model:] the density is given by Eq.~(\ref{Kingd}) with
central density $\rho _{\circ} = 6.0 \times10^4 \ \frac{{\rm M}_{\odot}}{{\rm pc}^3}$, 
core radius $r_c = 0.52 \ {\rm pc}$ and
${\rm M}_{{\rm tot}}^{{\rm King}} = 3.5 \times10^5 \,{\rm M}_{\odot}$
as already mentioned above.
\item[2. Inner isothermal model:] the density is given
 by Eq.~(\ref{regd})
with $r_{c} = 0.52 \ {\rm pc}$ and $\sigma = 10 \ {\rm km/s}$ as above,
but with $\rho _{\circ}$ chosen such that the total mass within the tidal
radius is the same as in the fitted King model, rather than determined via
Eq.~(\ref{corerad}).
\item[3. Singular model:] the mass density is described by 
the isothermal sphere, i.e. by Eq.~(\ref{singd})  
with the same $\sigma $ as mentioned above. The total mass of this model
within the tidal radius $r_t$ is ${\rm M}_{{\rm tot}}^{{\rm sing}}=2.8\times 10^6\,{\rm M}_{\odot}$.
\item[4. ${\bf \frac{1}{1+r^{2}}}$ - model:] the mass density is given by
\begin{equation} \rho (r) = \frac{\rho _{\circ}}{1 + 
\left( \frac{r}{r_{c}} \right)^{2}}\,
\end{equation}
with $r_c$ as above and $\rho _{\circ}$ chosen such that the total mass 
within the tidal radius is
the same as for the fitted King model.
\end{description}
The last three mass distributions are cutted discontinuously 
at the tidal radius.
Thus the integration
range for the optical depth and the lensing rate is only a sphere with 
a radius 
equal to the tidal radius.
For all models the velocity distribution follows
a Maxwell
distribution as in Eq.~(\ref{fdistr}) with $\sigma=10\,{\rm km/s}$.

The $\frac{1}{1+r^2}$-model is considered mainly because the integrations can be performed analytically.
Hence, we can easily compare the analytic result with the ones
obtained numerically for the other models.

Of course there exist more sophisticated models for globular clusters, e.g.
the (single or multimass) King-Michie models (King \cite{King:66}, 
Gunn \& Griffin \cite{Gunn:79}), that take into account the finite escape
velocity from the cluster, which naturally leads to a finite extension of
the cluster. However, for the study of the influence of different mass
distributions on the measured quantities, we consider the above mentioned 
models to be sufficient and, therefore, in this section, restrict our 
discussion to them. In addition, we remind on the well known scaling 
properties of the King model, which allow to derive the corresponding 
quantities for a different choice of parameters.

\subsection{Optical depth, lensing rate and mean event duration for SMC-47 Tuc}

We now discuss the different possibilities for microlensing in 
the system SMC-47 Tuc, i.e. events where the source and the lens are both located 
in 47 Tuc; the source is in the SMC and the lens in 47 Tuc; the source is
in 47 Tuc and the lens in the halo of the Milky Way, or finally the source resides 
in the SMC and the lens in the Milky Way. For SMC self-lensing we refer e.g. to 
the paper by Palanque-Delabrouille, Afonso, Albert et al. (\cite{npd}). 

At the end we also discuss the dependence on the mass function, which 
itself is independent of the lensing geometry. Since we hope to disentangle the 
different cases, we also calculate the differential rate
$\frac{d\Gamma}{dT}$. Throughout this part we will assume all lenses to have 
the same mass. The velocity distribution of the halo objects, as well as 
those of the cluster is taken to be a Maxwell function. As already
mentioned, we define 
the amplification threshold to be $A_{TH}=1.34$.

\subsubsection{Optical depth for source and deflecting mass in the cluster}

For completeness we discuss also this case, although, as we will see, its
contribution can be neglected for all practical purposes.

\begin{figure} 
\leavevmode
\epsfig{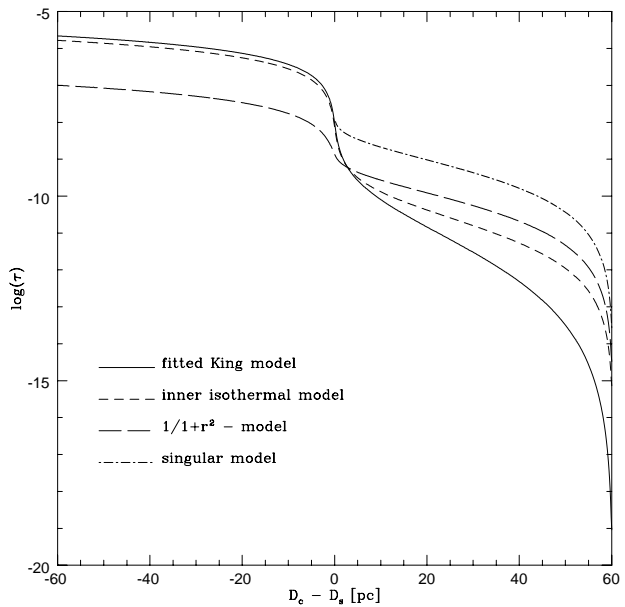}
\caption{The optical depth for the four different models as a function 
of the source position. All sources are located on the line of sight
from the observer to the cluster center, but at different radii relative
to the center. The optical depth of the singular model is only shown for
sources with $D_c-D>0$, since its mass density has a singularity at
$r=0$. 
For the plot we assume $f=1$, and the total mass of the models
as given in Sect.~3.
}
\label{difftaucc}
\end{figure}

For a pointlike source $\tau_{opt}$ is, according to 
Eq.~(\ref{pj3}), given by
\begin{equation} \label{taucc}
\tau_{opt} = \int_{x_b}^1 \frac{4 \pi G}{c^2} D^2 x (1-x) 
\rho _d(r(x)) \ dx
\end{equation}
where 
\begin{equation} \label{xh}
x_b = \frac{D_c - \sqrt{r_{t}^2 - b^2}}{D}\simeq 1 - \frac{r_t}{D} = 1 -
1.4 \times 10^{-2} \,.
\end{equation}
$D_c$ is the distance from the observer to the cluster center and $D$
the one to the source.
Hence, the integration is cut at the boundary of the cluster.
The optical depth for a source located in the center of the cluster 
($b=0$) is 
$\tau_{opt}=9.2 \times 10^{-9}
 \ \frac{M_{{\rm dark}}}{3.5 \times 10^5 \ M_{\odot}}$ 
for the  fitted King model.
 The results for the 4 different models 
 are shown in Fig.~\ref{difftaucc}. Of
course the isothermal-sphere model differs most, because its total mass
$M_{{\rm tot}}^{{\rm sing}} = 2.8 \times 10^6 M_{\odot}$ varies substantially from the
others.

Since the lower limit of integration is very close to $1$,
$\tau_{opt}$ can be approximated as follows
\begin{eqnarray*}
\tau_{opt} & \simeq & 
\frac{4 \pi G}{c^2} \int_0^{x_u} l \ \rho _d
(r(l)) \ dl \, .
 \end{eqnarray*}
Here $$x_u=\sqrt{r_t^2-b^2} - (D_c-D) $$
and $l=D-D_d$ is the distance from the lens (located at $D_d$) to the source. 
Hence, $\tau_{opt}$ is proportional to $D-D_d$ weighted with the mass density
in the globular cluster along the line of sight from the observer to the 
source.

\subsubsection{Lensing rate for source and deflecting mass in the cluster}

We assume the mass function to be independent of the position 
and all lenses are taken to have the same mass $\mu_{\circ}$ (in units of
$M_{\odot}$).
From Eq.~(\ref{fdist}) we find that the
transverse velocity distribution is given by 
\begin{equation}\label{meq12}
f_T(v_T) = 2\frac{v_T}{v_H^2}e^{-v_T^2/v^2_H}
\end{equation}
with $v_H^2=2 \sigma^2$.
If the source as well as the deflecting mass are in the cluster, the
event rate becomes
\begin{eqnarray} \label{Gammacc}
\Gamma & = & \frac{\sqrt{2 \pi} \sigma
r_E}{M_{\odot} \sqrt{\mu _{\circ}}} \ D \ 
\int_{x_b}^1 \rho _d(r(x)) \ \sqrt{x (1 - x)} dx\,,
\end{eqnarray}
with
$r(x) = \sqrt{b^2+(D_c-xD)^2}$
and $r_E$ defined to be 
\begin{equation}\label{myre}
r_E=\sqrt{\frac{4G}{c^2}M_{\odot}D}\,.
\end{equation}
We use the same geometry as in the previous subsection.
The result for the fitted King model for a source located at the center of
the cluster is
$\Gamma = \frac{6.6 \times 10^{-15}}{\sqrt{\mu_{\circ}}} \frac{M_{{\rm dark}}}
{3.5 \times 10^5 M_{\odot}} $ 1/s.

\begin{figure}
\leavevmode
\epsfig{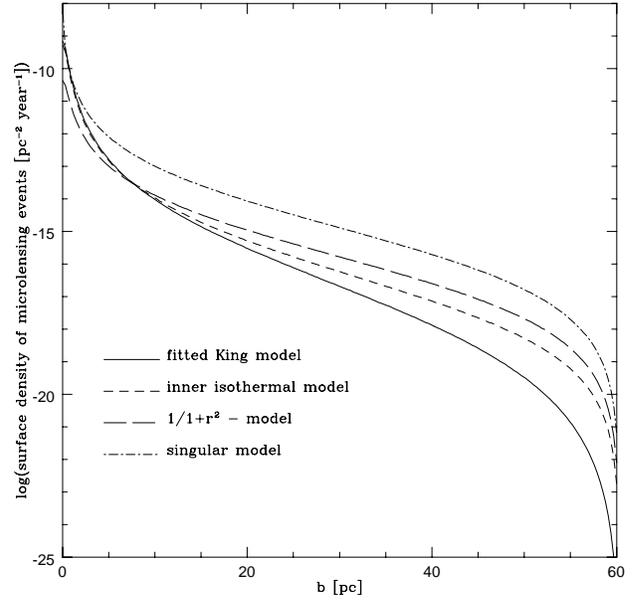}
\caption{The surface density of microlensing events is plotted as a function
of the projected radius $b$ for the four different models calculated in the
text, assuming $\mu_{\circ}=1$. For the plot we assume $f=1$, and the total
mass of the models
as given in Sect.~3.}
\label{diffintgam}
\end{figure}

In order to more easily compare between the microlensing rates for
different locations of the source and the lens, we introduce the quantity
$\tilde n(\alpha)$:
\begin{displaymath}
\tilde n(\alpha) \equiv \frac{\tilde N(\alpha)}{(1')^2} \ , 
\end{displaymath}
where $\tilde N(\alpha)$ is the number of microlensing events per unit
time in an area of
$(1')^2$ located at an angular distance $\alpha$ from the cluster center. 
For simplicity, we will give the rate $\tilde n(\alpha)$
in units of pc. In the plane 
perpendicular to the line of sight through the cluster
center,  $1'$ corresponds to 1.2 pc.
Hence, we can define a new quantity
$$ n\left(b=\alpha \frac{1.2\, {\rm pc}}{1'}\right) \equiv \tilde n(\alpha)
\frac{1}{(1.2\,
{\rm pc})^2} \ , $$ which is in units of ${\rm pc}^{-2}$ ${\rm unit\,\, time}^{-1}$.
We call $n(b)$ the surface density of microlensing events.
To calculate $n(b)$ we have to add up the lensing rates for all stars
located on the line of sight with impact parameter $b$ from the cluster
center. From Eq.~(\ref{Gammacc}), we see that the lensing rate $\Gamma$
depends on $D$ and on $b$, through $r(x)$ and $x_b$. Taking this into
account we get
\begin{equation} \label{microsd}
n(b) = \int_{D^i}
^{D^f}
 n_{{\rm star}}(\sqrt{b^{2} + (D_{c} -
D)^{2}}) 
 \ \Gamma (D\,,b) \ dD
\end{equation}
where in the ideal situation, which will lead to an upper bound for $n(b)$ 
we have $D^{f,i} = D_c \pm \sqrt{r_t^2 - b^2}$.
$n_{{\rm star}}(r)$ is the number density of stars (in units of ${\rm pc}^{-3}$) 
in the cluster at distance $r$ from the center
and $D$, $b$, $D_c$ are all in units of pc.
To describe the distribution of stars in the cluster, we assume that the
number density is proportional to the mass density $\rho_{{\rm King}}$ of the
fitted King model, since the mass surface density of this model is
proportional to the number surface density of stars in a cluster.
Thus $n_{{\rm star}}$ is given by
\begin{equation} \label{stardist}
n_{{\rm star}}(r) = \frac{N_{{\rm star}}}{M_{tot}^{{\rm King}}} \ \rho
_{{\rm King}}(r)
\end{equation}
with $N_{{\rm star}} = 4.6\times 10^5$ (Lang \cite{Lang:92}). 
With Eqs.(\ref{Gammacc}) and (\ref{microsd}) this leads to
\begin{eqnarray}
& & n(b) = 3.6 \times 10^{-19} \frac{N_{{\rm star}}}{M_{{\rm tot}}} 
\frac{1}{\sqrt{\mu_{\circ}}}
 \nonumber\\ 
& & 
\times \int_{D^i}^{D^f} \hspace{-0.3cm} D^{3/2} \rho _{{\rm King}} \left(\sqrt{b^2 + 
(D_c - D)^2} \right)
 \nonumber   \\ 
& & \times \int_{x_{b}(D)}^1 \hspace{-0.6cm} \sqrt{x (1 - x)}  
\rho_d \left(\sqrt{b^2 +
(D_c - x D)^2} \right) \ dD \ dx \,.\nonumber
\end{eqnarray}
The results for the four models are shown in Fig.~\ref{diffintgam}. 

The number of lensing events per year in the whole cluster is 
\begin{equation} \label{total}
\int_{0}^{r_{t}} 2 \pi \, n(b)b \, db
 = 2\times 10^{-2} \sqrt{\frac{1}{\mu_{\circ}}}\,
 f\, \frac{N_{{\rm star}}}{4.6 \times 10^{5}}  
\end{equation}
for the fitted King model.
On the other hand, the number of lensing events depends also on
the telescope resolution. Assuming a distribution
according to Eq.~(\ref{stardist}), we find that stars with a projected
radial distance smaller than $0.6\,$pc from the center cannot be resolved 
if the limiting
resolution of the telescope is $0.1"$, whereas if it is 
$1"$ even
stars with projected radius $b$ smaller than $7.5\,$pc from the center 
cannot be resolved. 
Hence, the number of expected lensing events per year 
using a telescope with a resolution of $0.1"$ is
\begin{displaymath}
\int_{0.6 {\rm pc}}^{r_t} 2 \pi \, n(b)b \, db = \,
\left\lbrace\begin{array}{ll}
1.1 \times 10^{-2}  \ \sqrt{\frac{1}{\mu_{\circ}}} \ \frac{M_{{\rm dark}}}{3.5
\times 10^5
 M_{\odot}} 
  \frac{N_
 {{\rm star}}}{4.6 \times 10^5}  & \quad  \\
 &     
	\\
2.4 \times 10^{-3}  \ \sqrt{\frac{1}{\mu_{\circ}}} \
\frac{M_{{\rm dark}}}{3.5 \times 10^5
 M_{\odot}}   
 \frac{N_
 {{\rm star}}}{4.6 \times 10^5} & \quad 
\end{array}\right.   
\end{displaymath}
and the one for a resolution of $1"$ is
\begin{displaymath}
\int_{7.5 {\rm pc}}^{r_t} 2 \pi \, n(b)b \, db = \,
\left\lbrace\begin{array}{ll}
1.4 \times 10^{-4}  \ \sqrt{\frac{1}{\mu_{\circ}}} \ \frac{M_{{\rm dark}}}{3.5
\times 10^5
 M_{\odot}} 
  \frac{N_
 {{\rm star}}}{4.6 \times 10^5}  & \quad \\
  & 
\\
2.3 \times 10^{-4}  \ \sqrt{\frac{1}{\mu_{\circ}}} \
\frac{M_{{\rm dark}}}{3.5 \times 10^5
 M_{\odot}}   
 \frac{N_
 {{\rm star}}}{4.6 \times 10^5}  & \quad 
\end{array}\right.   
\end{displaymath}
The first line is for the fitted King model, the second
for the $\frac{1}{1+r^2}$-model.

The distribution of the microlensing events as a function of their duration
T, using the transverse velocity
distribution as defined in Eq.~(\ref{meq12}) and the definition of $r_E$ in
Eq.~(\ref{myre}), is given by
\begin{eqnarray}\label{tdist}
& & \frac{d\Gamma}{dT} =  -\frac{2 D \mu_{\circ}}{M_{\odot}}
\frac{r_E^4}{\sigma ^2} \frac{1}{T^4}\nonumber \\
& & \times \int (x (1 -x))^2 \ \rho _d(x)
 \exp\left\lbrack-\frac{R_E^2(x)}{2 \sigma ^2 \ T^2}\right\rbrack \ dx. 
\end{eqnarray}
 
\begin{figure}
\begin{center}
\epsfig{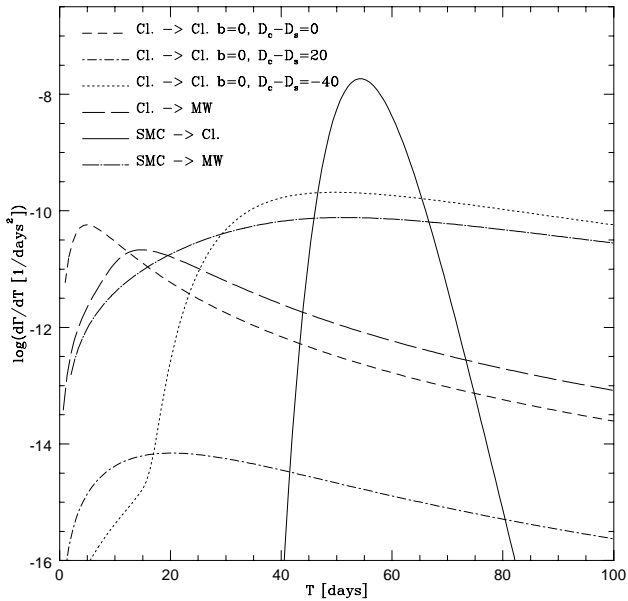}
\caption{The distribution of the lensing events with duration
$T=R_E/v_T$ for different geometries of the lens system.
For the calculations we used the fitted King model and made the assumption
that all lenses have the same mass $\mu_{\circ}=1$. The notation is to be
understood as follows: the first entry denotes the position of the source,
the second the one of the lens e.g. SMC$-\hspace{-0.2cm}>$Cl means a star 
of the SMC is
lensed by an object in the cluster. MW abbreviates Milky Way; $b,\,D_c$
and $D$ are defined within the text.
For the plot we assume $f=1$, and the total mass of the model
as given in Sect.~3.}
\label{figtdist}
\end{center}
\end{figure}

The result for the fitted King model assuming a lens mass $\mu_{\circ}=1$ is shown in
Fig.~\ref{figtdist}.

\subsubsection{Optical depth and lensing rate for the source in the SMC 
and the deflecting mass in the
cluster}

What changes in the calculation of $\tau_{opt}$ and $\Gamma$ 
is itemized below:
\begin{itemize}
\item the distance $D$ of the source is now about $63\,$kpc, corresponding to 
the distance of the SMC from the Sun; 
\item the $x$-integration goes from $x_{i} = 
(D_{c} - \sqrt{r_{t}^{2}-b^{2}})/D$, \\
 to $x_{f} = (D_{c} + \sqrt{r_{t}^{2}-b^{2}})/D$;
\item the velocity distribution gets shifted, because of
the motion of 47 Tuc relative to the SMC with a velocity of about $v_{d}$
$\cong$ $180\,$km/s. Hence, a velocity drift has to be 
introduced in the
Maxwell distribution, such that
\begin{equation} \label{fv}
f(\vec v) = \frac{1}{(2 \pi \sigma ^2)} 
\exp \left( - \frac{(v_x -v_d)^2 + v_y^2
+ v_z^2}{2 \sigma ^2} \right) \ ,
\end{equation}
where we assume the x-axis to be parallel to the velocity drift.
The resulting transverse velocity distribution is then given similarly to
Eq.~(\ref{meq12}).
With this, the integral over $dv_T$ in  Eq.~(\ref{l1}) yields
\begin{equation}
   \int_{0}^{\infty} v_{T} f(v_{T}) dv_{T} = v_d \simeq 180\,{\rm km/s}\,.
\end{equation}
\end{itemize}
Thus, the lensing rate and the optical depth become
\begin{eqnarray}
\tau_{opt} & = & \frac{4 \pi G}{c^2} D^2 \ \int_{x_i}^{x_f} x(1-x) \rho_d(r(x)) dx
 \ , \\
\Gamma & = & \frac{2 v_d r_E}{M_{\odot} \sqrt{\mu_{\circ}}} D \int_{x_i}^{x_f}
\ \sqrt{x(1-x)} \ \rho_d(r(x)) dx \nonumber\\
& = & \frac{5.1 \times 10^{-18}}{M_{\odot} \sqrt{\mu_{\circ}}} \left(\frac{D}{{\rm
pc}} \right)^{3/2} \
\hspace{-0.4cm}\int_{x_i}^{x_f}\hspace{-0.3cm} \sqrt{x (1-x)} \frac{\rho_d
(r(x))}{M_{\odot}/{\rm pc}^3} dx
\,.
\end{eqnarray}
For the extreme case of a star in the SMC lying on the line of sight going
through the center of the cluster, the optical
depth is $\tau_{opt} = 1.4 \times 10^{-4} \frac{M_{dark}}{3.5 \times 10^5 M_{\odot}}$ 
and the lensing rate $\Gamma = 2.0
 \times 10^{-11} \sqrt{\frac{1}{\mu_{\circ}}} \frac{M_{dark}}{3.5 \times 10^5 M_{\odot}}$ 1/s.

Since the tidal radius is smaller than 
about $200\,$pc, the quantity $x (1-x)$ as well as $\sqrt{x (1-x)}$ 
does not vary more 
than 10\% over the integration range,
and the variation of $\tau_{opt}$ as well as $\Gamma$ is directly 
proportional to the
surface density of the cluster. 
Assuming a tidal radius of $60\,$pc we find (with $\bar x$ an average value
for $x$)
\begin{eqnarray} 
& & \tau_{opt} = \bar x (1 - \bar x) \ \frac{4 \pi G}{c^{2}} \ D \ 
\Sigma(b)\,f \nonumber\\
& & = (2.32 \pm 0.04) \times 10^{-9} \left(\frac{D}{63\,{\rm kpc}} \right)\times \frac{\Sigma(b)}{
(M_{\odot}/{\rm pc}^2)}\,f \label{propor1}
\end{eqnarray}
\begin{eqnarray}
& & \Gamma(b) = \sqrt{\bar x (1- \bar x)} \ \frac{2 v_{d}
r_E}{M_{\odot}\sqrt{\mu_{\circ}}} 
 \ \Sigma(b)\,f 
                        \nonumber\\
& & =  (3.25 \pm 0.03) \times 10^{-16} 
\sqrt{\frac{D}{63\, {\rm kpc}}} \frac{\Sigma(b)}{M_{\odot}/{\rm pc}^2} \ 
\sqrt{\frac{1}{\mu_{\circ}}}\,f\,. \label{propor2} 
\end{eqnarray}
The mean event duration is $$\langle T \rangle = \frac{2}{\pi} 
\frac{\tau_{opt}}{\Gamma } = 52 \,\sqrt{\mu_{\circ}}\,\,\,{\rm days},$$
which is nearly independent of the angle between source and cluster center,
since both $\tau_{opt}$ and $\Gamma$ are proportional to $\Sigma(b)$ as given by 
Eqs.(\ref{propor1}) and (\ref{propor2}).

Of course the chance to find a lensing event in this case depends not
only on the optical depth, but also on the number of SMC stars in the
background of 47 Tuc.
Hesser, Harris, Vandenberg et al. (\cite{Hesser:87}) finds 
an average of $1000 \ {\rm stars}/30 ({\rm arcmin})^2$ with a magnitude in the
intervall of $0.0 < (B - V)<
0.8,\,\, 21 < V < 24$. 
In the following we 
assume a constant number of $50 \ {\rm stars}/({\rm arcmin})^{2}$,
although there is a gradient
in the density of the SMC stars across the whole 47 Tuc region
decreasing from the southeast to the northwest side. 

\begin{figure}
\begin{center}
\epsfig{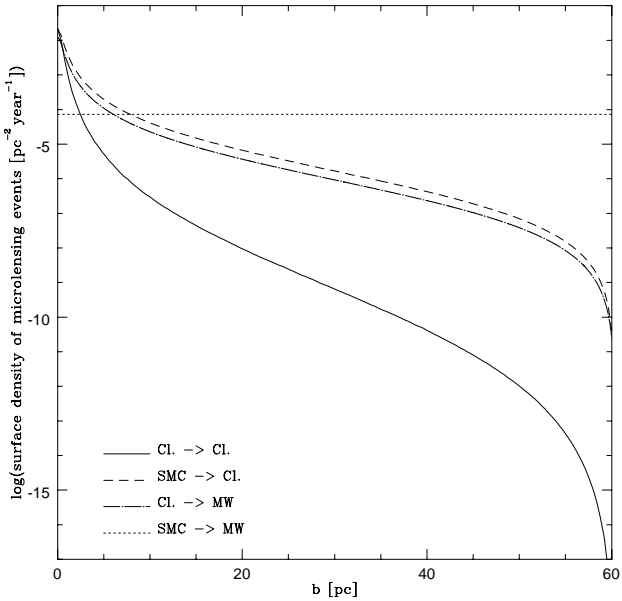}
\caption{The surface density of microlensing events is shown for the different
geometries of the lens system. For the calculations we used the fitted
King model and made the assumption, that all lenses have the same mass
$\mu_{\circ}=1$. For the abbreviations we refer to Fig.
\protect\ref{figtdist}.
For the plot we assume $f=1$, and the total mass of the models
as given in Sect.~3.}
\label{lensev}
\end{center}
\end{figure}

According to Eq.~(\ref{microsd}) we get 
with the above mentioned proportionality between $\Gamma$ and $\Sigma$
a surface density of microlensing events of
 \begin{equation}\label{eq40}
 n(b) = \sqrt{\bar x(1-\bar x)} \ \frac{2 v_d r_E}{M_{\odot}\sqrt{\mu_{\circ}}} \ \Sigma(b)
 \ \frac{n_{{\rm star}}}{(1.2 \,{\rm pc})^2}\,f \,,
\end{equation}
where $n_{{\rm star}}$ is the number of SMC stars per $(1')^2$.
$n(b)$ is shown in Fig.~\ref{lensev}.

The number
of events per year in the whole area of 47 Tuc is then given by
\begin{eqnarray}  \label{unknown}
  & \int_{0}^{r_t} 2\pi\,n(b) 
\, b db = & \nonumber \\
& = 0.12 \ \sqrt{\frac{1}{\mu_{\circ}}} \ \frac{M_{{\rm dark}}}{3.5 \times 10^5
 M_{\odot}}
\ \frac{n_
{star}}{50/(1.2 {\rm pc})^2} \ \frac{1}{{\rm year}} & 
\end{eqnarray}
for the fitted King model.
The expected number of events per year for an observation done with a
resolution of $0.1''$ is
\begin{displaymath}
\int_{0.6 {\rm pc}}^{r_t} 2 \pi\, n(b) \,b db = \,
\left\lbrace\begin{array}{ll}
1.0 \times 10^{-1}  \ \sqrt{\frac{1}{\mu_{\circ}}} \ \frac{M_{{\rm dark}}}{3.5
\times 10^5
 M_{\odot}} 
   \frac{n_
    {{\rm star}}}{50}   & \\
       &     
	\\
1.2 \times 10^{-1}  \ \sqrt{\frac{1}{\mu_{\circ}}} \
\frac{M_{{\rm dark}}}{3.5 \times 10^5
 M_{\odot}}   
 \frac{n_
   {{\rm star}}}{50} & 
\end{array}\right.   
\end{displaymath}
and for an observation with a resolution of $1''$
\begin{displaymath}
\int_{7.5 {\rm pc}}^{r_t} 2 \pi \,n(b) \,b db = \,
\left\lbrace\begin{array}{ll}
3.0 \times 10^{-2}  \ \sqrt{\frac{1}{\mu_{\circ}}} \ \frac{M_{{\rm dark}}}{3.5
\times 10^5
 M_{\odot}} 
   \frac{n_
    {{\rm star}}}{50}  & \\
       & 
	\\
1.0 \times 10^{-1}  \ \sqrt{\frac{1}{\mu_{\circ}}} \
\frac{M_{{\rm dark}}}{3.5 \times 10^5
 M_{\odot}}   
  \frac{n_
  {{\rm star}}}{50}\,.  & 
\end{array}\right.   
\end{displaymath}
The first value is for the fitted King-model and the second for the 
$\frac{1}{1+r^2}$-model. 

For comparison, if we insert a velocity $v_d 
\simeq 180\,{\rm km/s}=9.2 \times 10^{-3}$ arcsec/year for the cluster 
and a density of 50 SMC stars per $(1')^2$ in
the formula for the event rate per year, as given in the paper of Paczy\'nski
(\cite{Pac94}) 
one obtains
$f\times 1.5 \sqrt{\frac{1}{\mu_{\circ}}} \frac{1}{{\rm year}}$ 
for the rate in the whole area of the cluster,
$f\times 1.4 \sqrt{\frac{1}{\mu_{\circ}}} \frac{1}{{\rm year}}$ 
for the rate with a resolution of 0.1" and
$f \times 1.3 \sqrt{\frac{1}{\mu_{\circ}}} \frac{1}{{\rm year}}$
for the rate with a resolution of 1". The factor $f$ parametrizes the
fraction of dark matter $f=M_{{\rm dark}}/M_{{\rm tot}}$.
Paczy\'nski used in his calculations the total mass as given
by the singular isothermal sphere model, which gives $M_{{\rm tot}}=4.3
\times 10^6 M_{\odot}$.
The difference to our values is mainly due to the
different adopted mass surface densities. 
Since the total mass of the fitted King model is $M_{{\rm tot}}^{{\rm King}} = 3.5
\times 10^5 M_{\odot}$, this explains, with the help of Eq.~(\ref{unknown}), 
just the factor
12 between Paczy\'nski's results for the number of events
and ours for the fitted King model for the rate in the whole area of the
cluster. A multi-mass
model fitted to radial velocity and surface brightness observations of 47 Tuc 
leads to a total mass of $M_{tot}= 1.1 \times 10^6 M_{\odot}$ (Meylan
\cite{Meylan:89}), 
which is about
3 times bigger than the value for the fitted King model. In addition,
the fitted King model concentrates much more mass in the
center, whereas in the singular isothermal sphere model, there is still
a lot of mass in the outer regions of the cluster. From these
considerations, we see that the uncertainties are quite important and they
are mainly due to the poor knowledge of the dark matter content in globular
clusters. 

Paczy\'nski (\cite{Pac94}) proposed to measure the 
microlensing of background stars by MACHOs in globular clusters,
because in this way the mass of the lens can be determined more precisely 
than in the
lensing experiments under way.
Since the lens is in the cluster, it has the transverse velocity 
and the position of the cluster.
To estimate the inherent error of the method we adopt the values below,
where the transverse velocity $v_T$ is assumed to be precisely known up to the
cluster dispersion velocity and the distance up to the cluster
size: $v_{T} = 180\,{\rm km/s} \pm 10\,{\rm km/s}$,
$D_{d} = 4100\,{\rm pc} \pm 60\,{\rm pc}$
and $D = 63\,{\rm kpc} \pm 6\,{\rm kpc}$.
It follows then
an uncertainty in the mass determination of
$\left| \frac{\Delta m}{m} \right| \leq 0.33 \ .$
The uncertainty in the lens mass with the currently reported lensing
events is more like a factor 3 or even bigger 
(Jetzer \& Mass\'o \cite{JetzerM:94}, Jetzer \cite{Jet94}).

The distribution of the lensing events as a function of the duration
$T$ is:
\begin{eqnarray}
\frac{d\Gamma}{dT} & = & -\frac{2 D \mu_{\circ}}{\pi M_{\odot}}
\frac{r_E^4}{T^4} \frac{1}{\sigma ^2} \int_{x_E}^{x_b} dx
\int_0^{\pi} dy \ [x (1 - x)]^2 \ \rho _d(x) \nonumber \\
  &   & \quad \times \exp
\left(-\frac{v_d^2}{2 \sigma ^2} - \frac{R_E^2}{2 \sigma ^2 T^2} +
\frac{R_E \ v_d}{\sigma ^2 T} \cos y \right) \,, 
\end{eqnarray}
where $y$ is the angle between ${\bf v}_d$ and the projected MACHO velocity
in the plane perpendicular to the line of sight.
The numerical results, with the above mentioned values for $v_d$ and $\sigma$ 
are shown in Fig.~\protect\ref{figtdist}.

\subsubsection{Optical depth and lensing rate for the source in the cluster and 
the
deflecting mass in the halo of the Milky Way}

To calculate the optical depth and the lensing rate we set the position of
the source equal to the position of the cluster. In the reference frame,
where the origin is at the galactic center and the $x_1-x_2$ plane is the symmetry plane of the
galaxy, the coordinates of a lens, located on the
line of sight from the observer to the cluster center, are
$x_1  =  x D_c \cos b_T \cos l_T - R_{GC}$,
$x_2  =  x D_c \cos b_T \sin l_T$, and
$x_3  =  x D_c \sin b_T$, 
where $b_T = 44.89^o$ and $l_T = 305.9^o$ are the galactic latitude and
longitude of 47 Tuc.
With the number density distribution Eq.~(\ref{MWden}), we find the optical
depth (with $D=D_c$) to be $\tau_{opt} = 1.5 \times 10^{-8}$. 

Let's now look at the lensing rate. Since the velocity distribution in
the halo is assumed to be 
Maxwellian, the integration over $v_T$ in 
Eq.~(\ref{fdist}) leads to
\begin{equation}
\int v_T f(v_T) dv_T = \frac{v_H}{2} \sqrt{\pi}
\end{equation}
with $v_H\simeq 210\,{\rm km/s}$.
Thus, in a model where all lenses have the same mass, the lensing rate
becomes with Eq.~(\ref{l1}).
\begin{equation}
\Gamma = 4.7 \times 10^{-15}
\sqrt{\frac{1}{\mu_{\circ}}} \ 1/s\,.
\end{equation}

According to Eq.~(\ref{microsd}) the surface density of microlensing events
in this case is
\begin{equation}
n(b) = \Gamma \ \frac{N_{{\rm star}}}{M_{{\rm tot}}} \ \Sigma_{{\rm King}}(b) \ .
\end{equation}
$\frac{N_{{\rm star}}}{M_{{\rm tot}}} \Sigma_{{\rm King}}(b)$ is the number surface density 
of the stars in the cluster in units of ${\rm pc}^{-2}$.
  
The time scale calculated with $\Gamma$ and $\tau_{opt}$ as obtained above is
$$ <T> = \frac{2}{\pi} \frac{\tau_{opt}}{\Gamma} = 23 \ \sqrt{\mu_{\circ}} \
{\rm days}\,.
$$
For the distribution of the lensing events 
we obtain
\begin{eqnarray}
& & \frac{d\Gamma}{dT}=  -\frac{4 D \mu_{\circ}}{M_{\odot}} \frac{r_E^4}{v_H^2} 
\frac{1}{T^4}\nonumber \\
& &  \times \int_0^1 dx \rho _d(x) [x(1-x)]^2 \exp 
\ \left(-\frac{R_E^2}{v_H^2 T^2}
\right) 
\end{eqnarray}
with the MACHO mass density $\rho(x)$ as in Eqs.~(\ref{MWden}) and (\ref{norma}).
\begin{eqnarray}
& & \frac{d\Gamma}{dT}= 3.0\times 10^{-2} \frac{1}{T^4} \times \nonumber \\
& & \int_0^1 dx \rho(x) 
[x(1-x)]^2 \exp
\left(\llap{-}2.3\times 10^3 \frac{x (1 - x)}{T^2} \right) \frac{1}{{\rm days}^2}.
\nonumber \\
& & 
\end{eqnarray}
The numerical results with $v_H = 210 \ {\rm km/s}$ and $D = 4.1 \ {\rm
kpc}$ for $\mu _{\circ}=1$
are shown in Fig.~\ref{figtdist}
where the motion of the cluster relative to the
Milky Way halo is neglected. 

For a list of globular clusters which might be used as targets for a
systematic microlensing search, we refer to the papers of Gyuk \& Holder
(\cite{gyu}) and Rhoads \& Malhotra (\cite{rhoa}).

\subsection{Optical depth and lensing rate for a source in the SMC and the
deflecting mass in the halo of the Milky Way}

Since this is a standard case we restrict to the presentation of the 
results.
With a distance $D = 63\,$kpc
and the galactic coordinates of the SMC
$ l = 302.8^{\circ}$ $b = -44.3^{\circ} $,
we get for the optical depth and the lensing rate
\begin{eqnarray*}
\tau = 7.0 \times 10^{-7} \,\,\,{\rm and}\,\,\,
\Gamma = 6.6 \times 10^{-14} \sqrt{\frac{1}{\mu_{\circ}}} \ 1/s\,. & & 
\end{eqnarray*}
Therefore, the time scale is
$ <T> = 78 \ \sqrt{\mu_{\circ}}$ days. 
The result for $\frac{d\Gamma}{dT}$ is shown in Fig.~\ref{figtdist}.

\subsubsection{Dependence of the lensing rate $\Gamma$ on the mass
function }

Up to now, we assumed all lenses in the cluster to have the
same mass, which is, of course, rather unphysical. Therefore, we
discuss now the variation of $\Gamma$ with the
mass function $dn/d\mu$. 
However, we will still make the assumption that the
mass function does not depend on the position. To
estimate the variation of $\Gamma$ we choose a mass function of the
form
\begin{equation}\label{mfg}
\frac{dn_{\circ}}{d\mu } = C \mu ^{-(1+\gamma)}
\end{equation}
with $\gamma$ in the interval [-1,\,3] and the mass $m$ in the range
$m_a\leq m \leq m_b$.
With Eq.~(\ref{mfg}) the lensing rate $\Gamma$ becomes (with $u_{TH}$=1)
\begin{eqnarray}
\Gamma & = & \Gamma_{\odot} \frac{M_{\odot}}{\rho_{\circ}^{d}} \int \mu ^{1/2}
\frac{dn_{\circ}}{d\mu} d\mu \nonumber \\
 & = & \Gamma _{\odot} 
 \frac{1 - \gamma}{m_b^{1-\gamma} - m_a^{1-\gamma}}\times
 \frac{m_b^{1/2-\gamma} - m_a^{1/2-\gamma}}{1/2 - \gamma} \quad \,. 
\end{eqnarray}
$\Gamma_{\odot}$ is the lensing rate computed under the assumption that all
lenses have mass $\mu_{\circ}=1$. 

\begin{figure}
\leavevmode
\epsfig{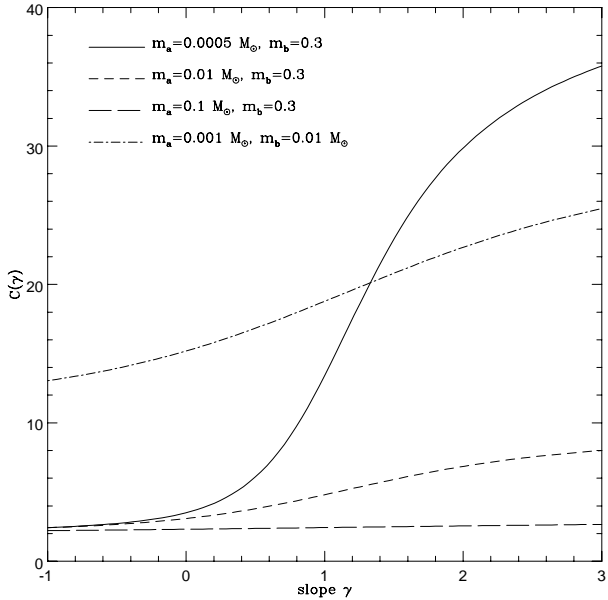}
\caption{$\Gamma/\Gamma_{\odot}$ is plotted as a funtion of the slope $\gamma$.
This corresponds to the dependence of the lensing rate on the slope 
of the mass function.
$\Gamma_{\odot}$ is the lensing
rate for a model where all MACHOs have the same mass 
 $\mu_{\circ}=1$}
\label{mfn}
\end{figure}

For the limiting cases $\gamma << 1/2$ and $\gamma >> 1$ this leads to a lensing
rate
\begin{equation}
\Gamma = \Gamma _{\odot} \left\{
\begin{array}{ll} 
		   \frac{1}{\sqrt{m_a}} & \quad \gamma >> 1 \\
		   \frac{1}{\sqrt{m_b}} & \quad \gamma << 1/2\,.
\end{array} \right.
\end{equation}
In Fig.~\ref{mfn} $\Gamma/\Gamma_{\odot}$ is shown for different 
upper and
lower limits of the MACHO mass. 

\section{Microlensing towards the bulge using dark objects in
globular clusters as lenses}

One encounters several problems by using globular cluster stars as sources. 
In fact, globular clusters have at most some 
$10^5$ stars that can be monitored. 
The angular size of a globular cluster is small and the stars are 
highly concentrated towards the center. Furthermore, most globular clusters 
are located towards the galactic center which restricts the number of objects 
being well suited for an exploration of the halo. Therefore, microlensing 
by globular clusters only leads to valuable 
results, if one is able to simultanously observe dozens of clusters with a 
very high resolution for several years. At present this seems to be a 
somewhat to demanding task, however this could be feasible in the near
future.

In this section we discuss the possibility to use dark objects in
globular clusters 
towards the bulge as lenses (see also Taillet, Longaretti \& Salati \cite{tail},
\cite{tail2}), in front of rich regions of the galactic
center or the spiral arms. 
There is a large sample of possible 
clusters which could be used for this purpose (see Table~\ref{mytab}). 
The core radius of the dark component of the cluster can be larger than the core
radius obtained by surface luminosity measurements.
Since the size of the cluster is much smaller than its
distance, it can be assumed to be equal for all its members, moreover, 
the velocity dispersion inside the cluster is also small, 
hence globular clusters are well suited for determining the lens mass.

\subsection{Estimate of optical depth and event rate}

In the first part of this section, we give some (conservative) estimates
for the optical depth and the event rate due to some globular clusters
towards the bulge by scaling the results obtained for 47 Tuc under the 
assumption that the clusters contain only little dark matter. 
As was derived in Eq.~(\protect\ref{eq40}), the optical depth
due to a globular 
cluster is proportional to the surface density of microlensing events 
and the pure geometrical 
quantity $\bar x (1-\bar x)$. Assuming $\Sigma(b)$ to be proportional to the 
total visual magnitude $M_V$, we can calculate the contribution to the 
optical depth due to the clusters (see Table~\protect\ref{mytab}). 
Since the galactic bar is not too extended, we fix the distance 
of the sources to be $8.5\,$kpc.

\begin{figure}
\leavevmode
    \epsfig{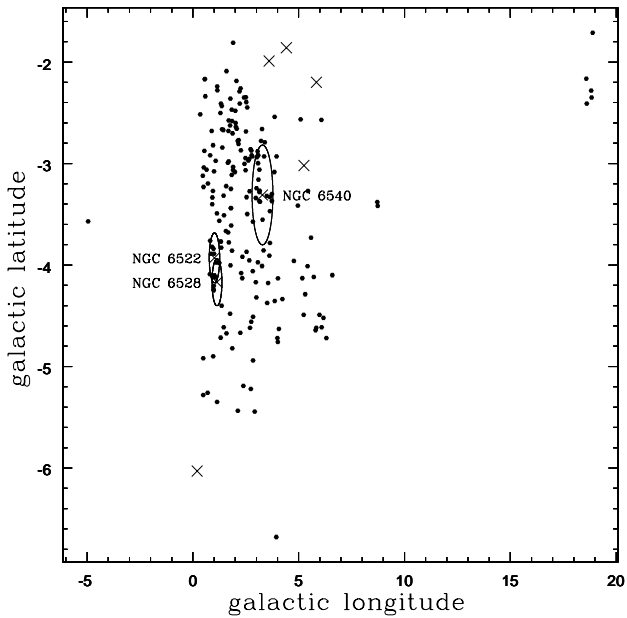}
    \caption{Position of 195 microlensing events towards the bulge
     	     in galactic coordinates (taken from the alert list
             of the MACHO collaboration and the corresponding list
             of the OGLE team). The crosses denote the position of
 	     some of the globular clusters reported in Table~\protect\ref{mytab},
	     which are located in Baades Window. Only the three clusters
             NGC 6522, NGC 6528 and NGC 6540 lie within the observation
             fields. The circles around the three clusters correspond to
             a radius of $30\,$pc around the cluster center.}
    \label{mlorte}
\end{figure}

\begin{table*}[h!tbp]
\begin{center}
\renewcommand{\arraystretch}{1.1}
\begin{tabular}{|l|l|c|c|c|c|}\hline
Object & Position $[l,\,b]$& $M_V$ & $\bar x$ & $\tau_{opt}\,\left\lbrack \frac{M_{{\rm dark}}}{3.5\times 10^5 M_{\odot}}\right\rbrack$
& N $\left\lbrack \sqrt{\frac{1}{\mu_{\circ}}} \frac{M_{{\rm dark}}}{3.5 \times 10^5
M_{\odot}}\frac{n_{star}}{50/(1.2 {\rm pc})^2} \frac{1}{{\rm year}} \right\rbrack$ \\ \hline
Terzan 1 & 357.57 1.00 & -3.3 & 0.765 & $2.1\times 10^{-7}$ & $1.1\times
10^{-5}$ \\ \hline
Terzan 5 & 3.81 1.67 & -7.9 & 0.941 & $4.7\times 10^{-6}$ & $5.5\times
10^{-4}$ \\ \hline
Terzan 6 & 358.57 -2.16 & -6.8 & 0.882 & $3.2\times 10^{-6}$ & $2.2\times
10^{-4}$ \\ \hline
UKS 1 & 5.12 0.76 & -6.2 & 0.882 & $1.8\times 10^{-6}$ & $1.3\times 10^{-4}$ \\ \hline
Terzan 9 & 3.60 -1.99 & -3.9 & 0.918 & $1.6\times 10^{-7}$ & $1.4\times
10^{-5}$ \\ \hline
Terzan 10 & 4.42 -1.86 & -7.8 & 0.988 & $9.0\times 10^{-7}$ & $4.3\times
10^{-4}$ \\ \hline
NGC 6522 & 1.02 -3.93 & -7.5 & 0.824 & $8.6\times 10^{-6}$ & $5.0\times
10^{-4}$ \\ \hline
NGC 6528 & 1.14 -4.17 & -6.7 & 0.871 & $3.2\times 10^{-6}$ & $2.1\times
10^{-4}$ \\ \hline
NGC 6540 & 3.29 -3.31 & -5.3 & 0.412 & $1.9\times 10^{-6}$ & $2.6\times
10^{-4}$ \\ \hline
NGC 6544 & 5.84 -2.20 & -6.5 & 0.294 & $4.9\times 10^{-6}$ & $1.5\times
10^{-3}$ \\ \hline
NGC 6553 & 5.25 -3.02 & -7.7 & 0.553 & $1.8\times 10^{-5}$ & $1.3\times
10^{-3}$ \\ \hline
NGC 6558 & 0.20 -6.03 & -6.1 & 0.753 & $3.0\times 10^{-6}$ & $1.6\times
10^{-4}$ \\ \hline
\end{tabular}
\caption{Optical depth and event rate due to MACHOs in some globular clusters 
for sources located towards the 
galactic center. The symbols are defined within the text. Cluster data is
adopted from Harris (\protect\cite{harr}).}
\label{mytab}
\end{center}
\end{table*}

Compared with the measured average optical depth towards the galactic center 
$\tau_{opt}\simeq 2.4\times 10^{-6}$ due to the disk and the bar itself 
(Alcock, Allsman, Alves et al. \cite{Alc97a}), we see 
from Table~\ref{mytab} that some of the globular clusters can give a very 
significant contribution to the total optical depth or even dominate it
along their line of sight, as 
in the case of NGC 6553, which lies close to the ideal distance 
of $\bar x=0.5$.

A rough estimate of
the microlensing event rate per year due to globular clusters 
towards Baades Window is obtained by properly rescaling 
Eq.~(\protect\ref{unknown}). We compute
the event rate in units of $\sqrt{\frac{1}{\mu_{\circ}}}\frac{M_{{\rm dark}}}{3.5 \times 10^5 M_{\odot}} \frac{n_
{star}}{50/(1.2 {\rm pc})^2} \frac{1}{{\rm year}}$ as in
Eq.~(\protect\ref{unknown}), in order to be able to easily compare between
the different clusters.
To that purpose one has to use the scaling factor $\eta$ given by:
\begin{equation}\label{scalf}
\eta=0.12\frac{v_d' r_E'}{v_d r_E}\frac{n'_{{\rm star}}}{n_{{\rm
star}}}\frac{(1.2\,{\rm pc})^2}{A}\frac{\int\Sigma'(b)\,db}{\int\Sigma
(b)\,db}\,,
\end{equation}
where the unprimed quantities belong to 47 Tuc and the primed to the cluster
under consideration. $A$ is the surface which corresponds to $(1')^2$ at the
distance of the cluster. For the calculation of the numbers, as given in 
Table~\ref{mytab}, we assumed $v_d'=30\,{\rm km/s}$ which is indeed a very
conservative value and, therefore, we will get a lower limit for the
eventrate. 

\subsection{Spatial distribution of microlensing events around NGC 6522,
NGC 6528 and NGC 6540}

In the following we present a rough analysis of the microlensing events
around the three globular clusters which lie within the observation
fields of MACHO and OGLE. 

Within a radius of $30\,$pc around NGC 6522 and 6528 we found
7 events for each of them (see Table~\ref{mytab2}). Since the projected areas of these two clusters
overlap, 4 events lie within the $30\,$pc circles around both clusters.
NGC 6540, which is nearer to us, hosts a total of 15 events within the 
$30\,$pc circle. At first sight, since the covered area is about four times
larger, this value seems to be lower than expected. However, NGC 6540
is about 8 times less bright than NGC 6522. 
Therefore, if the total luminosity roughly scales with the total mass
content, we expect NGC 6540 to contain less dark matter than NGC 6522.

Within $12\,$pc we found 3 events for NGC 6522, 2 events for NGC 6528 and
4 events for NGC 6540 (see Table~\ref{mytab2}). The event rate ratio in units of events per square
degree for the $12\,$pc circle and the $30\,$pc ring (excluding
the innermost $12\,$pc) are 99/25 for NGC 6522,
74/35 for NGC 6528 and 33/17 for NGC 6540, respectively. For all three clusters
the central region shows an increase in microlensing events. 
Due to the fact that the $30\,$pc
circles around NGC 6522 and NGC 6528 intersect, we can also calculate the event
rate in the overlapping region, which we find to be 70 events per square
degree. Within our poor statistics, this is what one expects for a line
of sight crossing twice the region of influence of a globular cluster. It
would also give a first hint, that globular clusters can indeed have a
population of dark objects that reaches out as far as $30\,$pc.

Of course, there is now the problem how to distinguish between the events due
to MACHOs located in the globular cluster and those due to MACHOs in the
disk or bulge, which will define our "background". Since this cannot be
decided for a single event, we assume that the events which lie in the ring
from $12$ to $30\,$pc are due to MACHOs in the disk or bulge. This way we
certainly overestimate the "background". Moreover, the common events 
within $30\,$pc around both NGC 6522 and NGC 6528 were counted as
"background" events for both clusters. This way leading also to a higher background
rate. The so estimated event rate per area is then subtracted from the value
in the inner $12\,$pc region. The leftover events should be due to
MACHOs located in the globular cluster.
We find the following values (in parenthesis the "background"):
2 (1) hence a total of 3 events for NGC 6522, 1 (1) event for NGC 6528
and 2 (2) events for NGC 6540. We see that the number of observed events in
the inner $12\,$pc is roughly twice as high as one would get due to MACHOs
in the disk or bulge alone. By assuming that all the events in
the ring from $12$ to $30\,$pc are due to MACHOs in the disk or bulge, we
have certainly underestimated the contribution from the globular cluster,
since some of these events might also be associated with the cluster.

Of course, we must discuss the shortcomings of our analysis. We 
tacitly assumed that the product of total observation time and background
star density is the same for the $12$ and $30\,$pc regions for a given
cluster. This
should at least be well fullfilled for the relatively small regions
around NGC 6522 and NGC 6528. For NGC 6522
and NGC 6528 we added MACHO and OGLE data, hence we use a different
normalisation for them than for NGC 6540. In addition, we did not
take into account the different efficiencies. Moreover, our evaluation 
is based upon a very poor statistics, and thus one
must take the above results with all the necessary caution. 
However, since all our estimates were performed very conservatively and
for all three clusters we get the same behaviour and since also the overlapping
region of NGC 6522 and NGC 6528 shows an increase in microlensing events, we
think it's fair to state, that globular clusters can --within some pc-- 
at least double the optical depth and also the event rate due to MACHOs
in the disk and the bulge; the latter quantities being $\tau \simeq
2.4\times 10^{-6}$ (Alcock, Allsman, Alves et al. \cite{Alc97a}) and $\Gamma\simeq 1.3\times 10^{-12}s^{-1}$ for a
typical mass of $0.1\,M_{\odot}$. 

Stars in globular clusters can also act as sources for microlensing,
however, as discussed in Sect.~3.2.4 their contribution, unless for
the very central region of $\sim 1-2\,$pc, is very small as compared with
lensing due to sources in the bulge. For NGC 6522 and NGC 6528 there is
a small probability that the source lies in NGC 6528 and the lens in NGC 6522.
Knowing the distances and the relative motion of the two clusters this
system might yield the most accurate mass determination for a lensing
object.

It is interesting to note, that the above mentioned event rate ratios scale
with the total luminosity i.e. the brightest cluster NGC 6522 shows also 
the largest increase of events towards the center.

For the mean event duration as given
in Table~\ref{mytab2}, we calculate the typical mass of a MACHO for
$v_d=30\,$km/s and $v_d=180\,$km/s. Obviously, one should consider
only the events due to MACHOs in the cluster. However, since it is
not possible to distinguish them, we have just taken the average value as 
a first approximation. This should not be far from the true value
(as can be seen by inspection of Table~\ref{mytab2}). The results are 
given in Table~\ref{mytab3}. We see that depending on $v_d$ the MACHOs 
can be either Jupiter type objects, brown dwarfs, M-stars or even
white dwarfs.

\begin{table*}[h!tbp]
\begin{center}
\renewcommand{\arraystretch}{1.1}
\begin{tabular}{|l|l|c|c|c|c|c|c|}\hline
Object & Position $[l,\,b]$& $M_V$ & within $12\,$pc& within $30\,$pc
&  Duration [days] & Mean [days] & Events/degree$^{2}$ \\ \hline
NGC 6522 & 1.02 -3.93 & -7.5 & OGLE 5 & & 12.4 & & \\
 & & & 97-37$^*$ & & 12 & & \\
 & & & 97-68 & & 6 & & \\ 
 & & & & & & 10.1 & 99 \\
 & & & & OGLE 1$^*$ & 25.9 & & \\
 & & & & OGLE 2 & 45 & & \\
 & & & & OGLE 4$^*$ & 14 & & \\
 & & & & 97-14$^*$ & 21.5 & & \\
 & & & & & & 26.6 & 25 \\
\hline
NGC 6528 & 1.14 -4.17 & -6.7 & OGLE 1$^*$ & & 25.9 & & \\
 & & & 97-14$^*$ & & 21.5 & & \\
 & & & & & & 23.7 & 74 \\
 & & & & OGLE 3 & 10.7 & & \\
 & & & & OGLE 4$^*$ & 14 & & \\
 & & & & OGLE 10 & 61.1 & & \\
 & & & & 97-37$^*$ & 12 & & \\
 & & & & 95-11 & 30.5 & & \\
 & & & & & & 22 & 35 \\
\hline
NGC 6540 & 3.29 -3.31 & -5.3 & 96-6, $d<6\,$pc & & 19.5 & & 33 \\
 & & & 96-17 & & 16.5 & & \\
 & & & 95-29 & & 12 & & \\
 & & & 95-40 & & 8.5 & & \\
 & & & & & & 14 & 33 \\
 & & & & 96-14, $d<18\,$pc& 12.5 & & \\
 & & & & 95-26,  $d<18\,$pc& 19.5 & & \\
 & & & & 95-31, $d<18\,$pc& 16 & & 20 \\
 & & & & 97-2, $d<24\,$pc& 23.5 & & \\
 & & & & 97-10, $d<24\,$pc& 8.5 & & \\
 & & & & 97-24, $d<24\,$pc& 5 & & \\
 & & & & 95-36, $d<24\,$pc& 12 & & 20 \\
 & & & & 98-1 & 97.5 & & \\
 & & & & 97-58 & 26.5 & & \\
 & & & & 96-1 & 76 & & \\
 & & & & 95-30 & 33.5 & & 15 \\
 & & & & & & 30 & 17 \\
\hline
\end{tabular}
\caption{Microlensing events within a radial distance of
$30\,$pc around NGC 6522, NGC 6528 and NGC 6540. Values marked with an
asterisk
lie within $30\,$pc of both NGC 6522 and NGC 6528. For
NGC 6540 we also give the finer binning as used for Fig.~\ref{rate6540}. 
Data is taken from the alert list of the MACHO collaboration and the 
event list of the OGLE team. The event duration follows the OGLE 
convention i.e. $T=R_E/v_T$, which is half the value as reported in the MACHO alert list.}
\label{mytab2}
\end{center}
\end{table*}

\begin{table}[h!tbp]
\begin{center}
\renewcommand{\arraystretch}{1.1}
\begin{tabular}{|l|l|l|}\hline
Object & $\langle M\rangle$  $d\leq 12\,$pc & $\langle M\rangle$  $12<d\leq 30\,$pc \\ \hline
NGC 6522 & 0.003$\,M_{\odot}$ & 0.021$\,M_{\odot}$ \\
 & 0.11$\,M_{\odot}$ & 0.75$\,M_{\odot}$ \\ \hline
NGC 6528 & 0.021$\,M_{\odot}$ & 0.018$\,M_{\odot}$ \\
 & 0.77$\,M_{\odot}$ & 0.65$\,M_{\odot}$ \\ \hline
NGC 6540 & 0.003$\,M_{\odot}$ & 0.016$\,M_{\odot}$ \\
 & 0.12$\,M_{\odot}$ & 0.57$\,M_{\odot}$ \\ \hline
\end{tabular}
\caption{Typical masses for the microlensing events around NGC 6522, NGC
6528 and NGC 6540. The upper value corresponds to $v_d=30\,$km/s, the
lower to $v_d=180\,$km/s. $d$ denotes the distance from the cluster center. 
}
\label{mytab3}
\end{center}
\end{table}

\begin{figure}
\leavevmode
    \epsfig{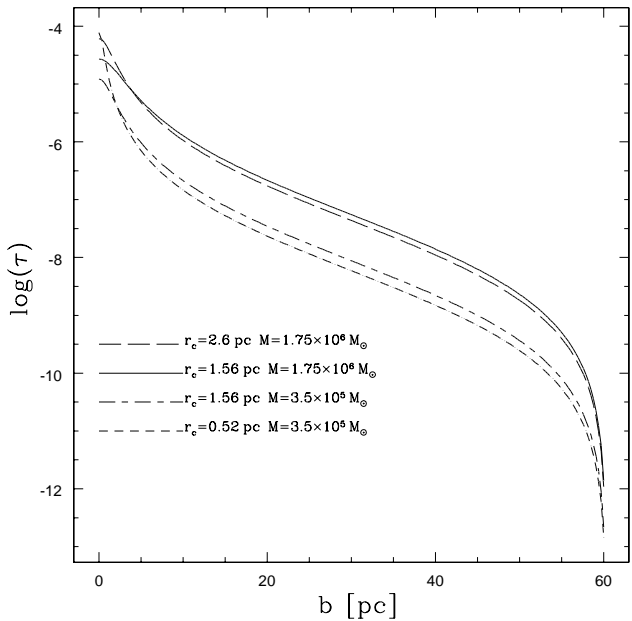}
    \caption{The optical depth for the four different King models
             as described within the text.
             }                       
    \label{taumult}
\end{figure}

\begin{figure}
\leavevmode
    \epsfig{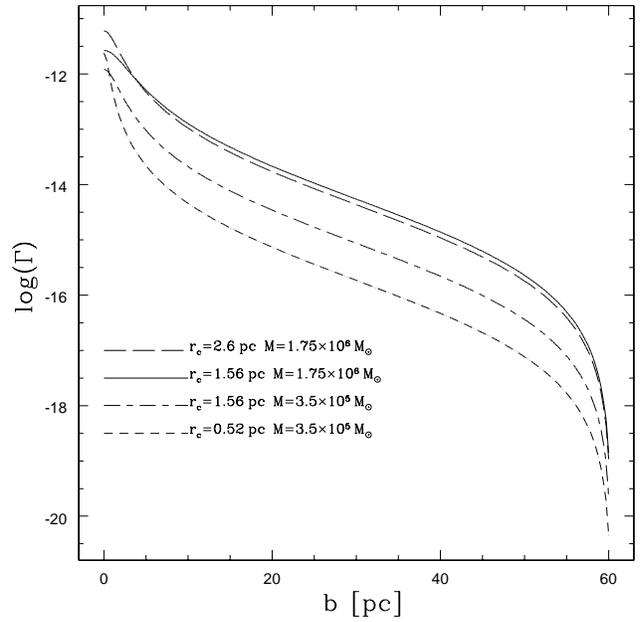}
    \caption{The microlensing event rate for the four different King models 
             as described within the text.
             }                       
    \label{gammamult}
\end{figure}

Below we compute the optical depth and the event
rate for four different King models as a function of the
distance from the cluster center in the lens plane. 
The parameters are tuned such
that the average of the values in the interval from $1$ to $12\,$pc
roughly corresponds to the above mentioned values for the optical depth 
$\tau$ and the event rate $\Gamma$. The tidal radius was assumed to
be $r_t=60.3\,$pc, the distance to the lens is set to be $3.5\,$kpc 
and the one to the source $8.5\,$kpc. For the calculation of the event
rate we again take the rather low value $v_d=30\,$km/s. Since we are not
able to reproduce the mass function, we rather study a bi-mass model, hence
the cluster consists of a heavy component (component 1) and a light one
described by one of the other components as defined below. 
\begin{description}
\item[Component 1:] the density is given by Eq.~(\ref{Kingd}) with
central density $\rho _{\circ} = 6.0 \times10^4 \ \frac{{\rm
M}_{\odot}}{{\rm pc^3}}$
core radius $r_c = 0.52 \ {\rm pc}$ and
${\rm M}_{1}^{{\rm King}} = 3.5 \times10^5 \,{\rm M}_{\odot}$. For the
calculation of the event rate we assumed a typical mass 
${\rm M}={\rm M}_{\odot}$.
\item[Component 2:] as Component 1, but with a core radius $r_c = 1.56 \ {\rm pc}$
and a typical mass of ${\rm M}=0.1\,{\rm M}_{\odot}$. The total mass of this
component is again $3.5 \times10^5 \,{\rm M}_{\odot}$.
\item[Component 3:] as Component 2, but with a total mass of the component 
of $1.75\times10^6
\,{\rm M}_{\odot}$
\item[Component 4:] as Component 3, but with a core radius $r_c = 2.6 \
{\rm pc}$
\end{description}
We find that a population of low mass objects as described by a King model 
with a total mass of $1.75\times10^6\,{\rm M}_{\odot}$
can lead to the desired enhancement of the optical depth and the lensing 
event rate up to distances of $\sim 12\,$pc from the cluster center 
(see Figs.~\protect\ref{taumult} and \protect\ref{gammamult}).
Of course, the models and mass values given above have to be taken as an illustration,
nevertheless it is clear that the rather high observed microlensing 
rate of the clusters imply a substantial dark matter component.
Although there is a large inherent uncertainty in the event rate due
to the poor knowledge of $v_d$, it is important to note, that the
required optical depth and event rate cannot be due to the 
heavy component alone.

\begin{figure}
\leavevmode
    \epsfig{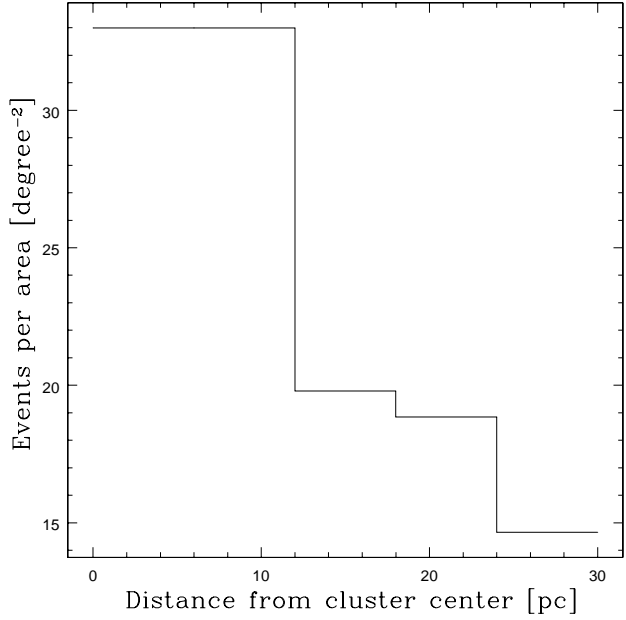}
    \caption{The microlensing event rate per area as a function of the radial
             distance from the cluster center for NGC 6540.
             }
    \label{rate6540}
\end{figure}

\section{Summary}

We discussed in detail microlensing by globular clusters. 
47 Tuc was taken as an example for which we performed the calculation
of the optical depth, the microlensing event rate and the average
lensing duration for all possible geometries of the 
system SMC-47 Tuc-Milky Way. In addition, we studied the dependence of
these parameters on the mass function. 

We have seen that for the case,
where the source is a star in the SMC and the lens is a MACHO in
47 Tuc, one can expect an observable eventrate of $\sim 0.1-1$ per year.
However, this result depends crucially on the total amount of dark
matter and its distribution in the cluster, which both are not well known 
at present.

We then applied these results to study microlensing by globular clusters
towards the galactic center, where locally the optical depth can be
dominated by dark matter inside clusters. However,
since globular clusters are very localized objects, 
the expected number of events as obtained by these scaling arguments is small.
A larger event rate is expected, if the average
MACHO mass inside the cluster is well below one solar mass, the total amount
of dark matter is larger than $3.5 \times 10^5 M_{\odot}$ or the density of
observable stars behind the cluster is significantly higher than the assumed
value of 50 stars per $(1.2 \,{\rm pc})^2$. 
Indeed, analysing the event distribution around the three clusters inside the
observation fields of MACHO and OGLE, we find an increase of the 
microlensing event rate by at least
a factor of about 2, as compared to that expected for MACHOs located in
the disk and the bulge. This increase suggests the presence of a substantial
amount of dark matter in form of light objects such as brown dwarfs.

Given this promising preliminary results it is important
to systematically analyse future lensing data as a function
of the position around the mentioned globular clusters. 
In fact, having few more events at disposal will already be very helpful to
draw more firm conclusions and get better limits on the content of dark matter
in globular clusters.

In addition, we propose to
favour observation fields around globular clusters for future campaigns.
In particular NGC 6553 is a very promising candidate for lensing by dark
objects in globular clusters, since its luminosity is high and also its
distance is such that the tidal radius of the cluster
corresponds to a relatively large angular size. Moreover, it would be
important to have more precise knowledge of $v_d$, which is at
present one of the main sources of uncertainty.

\begin{acknowledgements}
This work is partially supported by the Swiss National Science Foundation.
It is a pleasure to thank the referee for his helpful and
encouraging remarks.
\end{acknowledgements}

\end{document}